\documentclass[%
 preprint,
 twocolumn,
 superscriptaddress,
 amsmath,amssymb,
 aps,
 prl
]{revtex4-1}

\usepackage{graphicx}
\usepackage{dcolumn}
\usepackage{bm}

\begin{document}

\preprint{APS/123-QED}

\title{Low-Scaling Algorithm for Nudged Elastic Band Calculations Using a Surrogate Machine Learning Model}
\author{Jos\'e A. Garrido Torres} %
\author{Paul C. Jennings}%
\author{Martin H. Hansen}%
\author{Jacob R. Boes}%
\affiliation{
Stanford University, Department of Chemical Engineering, Stanford, California 94305, USA
}
\affiliation{
SUNCAT Center for Interface Science and Catalysis, Stanford Linear Accelerator Center, Menlo Park, California 94025, USA.
}
\author{Thomas Bligaard}%
 \email{bligaard@stanford.edu}
\affiliation{
SUNCAT Center for Interface Science and Catalysis, Stanford Linear Accelerator Center, Menlo Park, California 94025, USA.
}

\date{\today}

\begin{abstract}
We present the incorporation of a surrogate Gaussian Process Regression (GPR) atomistic model to greatly accelerate the rate of convergence of classical Nudged Elastic Band (NEB) calculations.
In our surrogate model approach, the cost of converging the elastic band no longer scales with the number of moving images on the path.
This provides a far more efficient and robust transition state search. 
In contrast to a conventional NEB calculation, the algorithm presented here eliminates any need for manipulating the number of images to obtain a converged result.
This is achieved by inventing a new convergence criteria that exploits the probabilistic nature of the GPR to use uncertainty estimates of all images in combination with the force of the transition state in the analytic potential.
Our method is an order of magnitude faster in terms of function evaluations than the conventional NEB method with no accuracy loss for the converged energy barrier values.

\end{abstract}

\pacs{Valid PACS appear here}
\maketitle


The Nudged Elastic Band (NEB) algorithm is the most popular method for calculating transition states in chemical systems \cite{jonsson1998nudged, henkelman2000improved, henkelman2000climbing}.
This algorithm is used to find minimum energy pathways (MEP) for the transition between reactants and products, identifying the energy associated with the barrier separating these two states.
Many variants of the NEB algorithm have been proposed in the last two decades \cite{henkelman2000climbing, maragakis2002adaptive, chu2003super, trygubenko2004doubly, peters2004growing, gonzalez2006searching, sheppard2012generalized, kolsbjerg2016automated}.
All of these algorithms rely on an elastic band consisting of interpolated images of the atomic structure, known as moving images.
The images are hooked by a spring constant and their positions are optimized by following the gradient of the potential energy surface (PES) while obeying the forces imposed by these springs.
A climbing image (CI), without spring forces and an added force traveling up the gradient along the tangent of the path, can also be included in order ensure the highest energy point is included in the band \cite{henkelman2000climbing}.
The optimization of the path is performed through an iterative process in which all the images are moved and evaluated in each iteration.
The coupled iterative nature of the process is very costly, requiring several hundred function calls for the forces even for systems containing few images and degrees of freedom, e.g. describing a single particle diffusion with 10 images.

Further, force evaluations can be computationally very expensive for the first-principle electronic structure calculations.
For this purpose, there has been significant work done to build machine learning (ML) surrogate models for atomistic systems \cite{khorshidi2016amp, peterson2016acceleration, koistinen2017nudged, jorgensen2018exploration, chen2018atomic, 2018arXiv180808588G}.
These methods function by producing a surrogate model of the PES, which closely approximates the analytic potential in the region of interest, significantly reducing the number of necessary function calls to achieve convergence.
Among all of these models, the critical steps are: (1) moving the atomic positions along the surrogate PES using traditional algebraic or derivative-based solvers, (2) evaluating analytically the forces at the new positions and (3) updating the model with the evaluated point(s) in order to improve the predicting capabilities of the surrogate model.
This iterative process is performed until convergence is reached.
The premise underlying this protocol is that the optimization cost of the PES surrogate is essentially negligible compared to the cost of an electronic structure calculation.

The aforementioned strategy has served to accelerate NEB calculations using neural networks (NN) as proposed by Peterson \textit{et al.} \cite{peterson2016acceleration} and using GPR by J\'onsson \textit{et al.} \cite{koistinen2017nudged}.
Both approaches have demonstrated the ability to reduce the high computational cost of the classical NEB methods. 
However, even in these cases, all moving images must be evaluated at least once to ensure that the convergence criteria has been satisfied.
To the best of our knowledge this also holds true for the other NEB algorithms proposed to date. 

One of the main advantages of using GPR is that, as a probabilistic model, the uncertainty estimate for the predictions can be quantified.
In this letter, we demonstrate that the efficiency of the current NEB algorithms can be substantially improved by choosing an acquisition function that optimally utilizes the prediction obtained by the GPR model, i.e. the Gaussian posterior distribution.

Following these principles, we also propose an algorithm that uses the GPR estimates to define a convergence criteria which is independent of the number of NEB images, therefore solving one the major problems of the previous classical and machine learning NEB methods.
This algorithm is implemented in CatLearn \cite{suncat2018catlearn}, which is an open-source Python package for machine learning applications specific to atomic systems.
This is, by design, built to interface with the Atomistic Simulation Environment (ASE) \cite{larsen2017atomic} and therefore can be easily interfaced with the majority of the electronic-structure calculators, such as CASTEP \cite{segall2002first}, GPAW \cite{enkovaara2010electronic}, Quantum Espresso \cite{giannozzi2009quantum}, SIESTA \cite{soler2002siesta}, and VASP \cite{kresse1996efficient, kresse1999ultrasoft}.

Our GPR model considers the positions of the atoms as the descriptors \textbf{X}~=~[$\textbf{x}_1$,~\dots~, $\textbf{x}_N$] and is trained with their corresponding energies ($\textbf{e}$) and first derivative observations ($\boldsymbol\delta_i$), combining both observations into a vector \textbf{y}~=~[\textbf{e} $\boldsymbol\delta_1$~$\dots$ {$\boldsymbol\delta_N$}].

The predicted function is \textit{a priori} defined as the Gaussian process:
\begin{equation}
f(x) \sim \mathcal{GP}(P(\textbf{x}), k(\textbf{x},~\textbf{x}')), 	
\end{equation}
where $k(\textbf{x},~\textbf{x}')$ is the kernel (covariance function) and $P(\textbf{x})$ is the prior function. In our model, we chose a constant prior of the form $P(\textbf{x})=(max(\textbf{e}), 0)$ and the square exponential kernel (SE),
\begin{equation}
	k(\textbf{x},~\textbf{x}') = \sigma{_f^2} ~exp\Bigg[ \frac{1}{2} \sum_{m=1}^d \frac{{(x_m - x_m^{'})}^2}{l_m^2} \Bigg],
\end{equation}
with ${l}_m$ and $\sigma{_f}$ being the characteristic length scale for each predictor and the signal standard deviation parameters, respectively.

When incorporating first derivative observations to the GP, the covariance matrix takes the form 
\[ \textbf{K}(\textbf{x}) = \left( \begin{array}{cc}
\textbf{K}(\textbf{x}, \textbf{x}) & \textbf{K}_{gd}(\textbf{x}, \textbf{x})  \\
\textbf{K}_{gd}(\textbf{x}, \textbf{x})^\top & \textbf{K}_{dd}(\textbf{x}, \textbf{x})  \end{array} \right),\]
with elements of the block matrix being the covariance between the coordinates (\textbf{K}(\textbf{x}, $\textbf{x})$), and partial derivatives of the covariance with respect to the first coordinate ($\textbf{K}_{gd}(\textbf{x}, \textbf{x})$), second coordinate ($\textbf{K}_{gd}(\textbf{x}, \textbf{x})^\top)$, and the first and second set of coordinates ($\textbf{K}_{dd}(\textbf{x}, \textbf{x})$).

Our dataset is defined as $\mathcal{D}$~=~$\big\{$$\big\{$$\textbf{x}_n$, $\textbf{e}_n$, \textbf{$\boldsymbol \delta_N$}, 
$\boldsymbol \theta$ \big\}\big\}$_{n=1}^N$, where $\boldsymbol \theta$ contains the set of hyperparameters of the model.
The predicted mean and variance of the GP are given by
\begin{equation}
	\label{eq:mean}
	\mathbb{E}[f(\textbf{x})|\mathcal{D}]=\textbf{k}(\textbf{x}) [\textbf{K}(\textbf{x}) + \sigma_n^2 \textbf{I}]^{-1} \textbf{y}
\end{equation}
and
\begin{equation}
	\label{eq:variance}
	\mathbb{V}[f(\textbf{x})|\mathcal{D}]=k(\textbf{x, x}) - \textbf{k}(\textbf{x})[\textbf{K}(\textbf{x}) + \sigma_n^2 \textbf{I}]^{-1} \textbf{k}(\textbf{x}),
\end{equation}
respectively, where $\textbf{I}$ is the identity matrix and $\sigma_n^2$ is a regularization parameter. The predicted mean (Eq. \ref{eq:mean}) provides the prediction of the energy for a given position whilst the predicted variance (Eq. \ref{eq:variance}) offers an estimate of the uncertainty of the same process.
In our model, the $\sigma_f$ parameter is kept fixed whilst the length-scale ($l$) is optimized isotropically.
The regularization term ($\sigma_n$) is added to the diagonal of the covariance matrix and is separately optimized for the elements involving the kernel function and the derivative terms of the kernels. 
This is relevant since the energies and the forces may have different sources of error.\\

\begin{figure}[!h]
\includegraphics[scale=0.85]{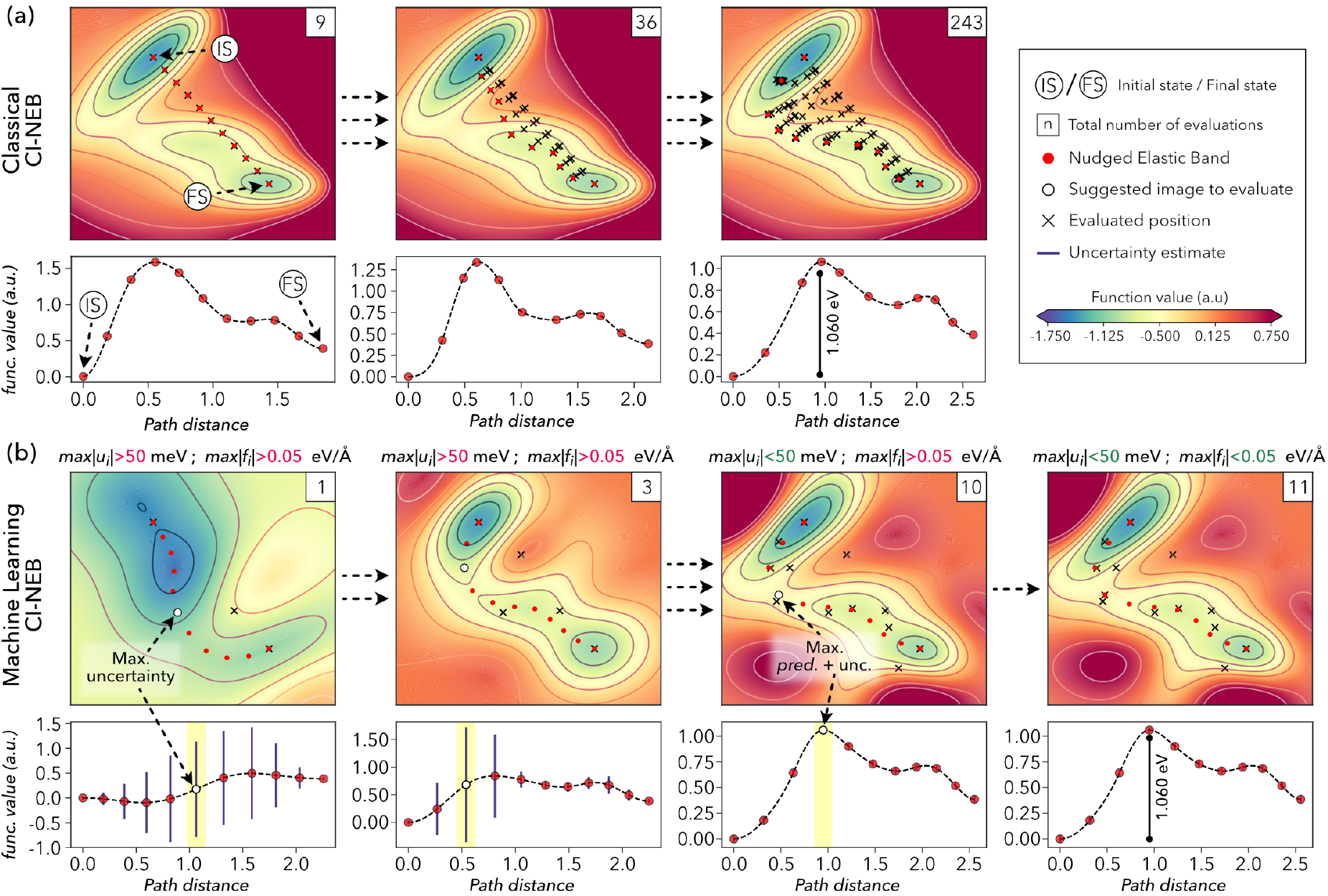}
\caption{Comparison between the (a) classical CI-NEB and (b) machine learning NEB (ML-NEB) methods. The performance of both algorithms is illustrated in the two-dimensional M\"uller-Brown PES. Predicted MEP is included at the bottom of each PES to show the evolution of the energy profile for the elastic band with respect to the number of function calls.}
\label{Fig1}
\end{figure}

A comparison between the classical NEB and our machine learning accelerated (ML-NEB) methods on the two-dimensional M\"uller-Brown potential is shown in Figure \ref{Fig1}. 
In this example, we used 9 moving images to describe the transition from the initial state (IS) to the final state (FS).
In the classical NEB method, final convergence is achieved when the maximum forces of the structure of the $i$th NEB image (max$|F^{NEB}_{i}|$) perpendicular to the path are below the convergence criteria. 
This convergence criterion (max$|$$F^{NEB}_{i}|$$<$0.05~eV/\AA) is satisfied after 243 force calls with an energy barrier of 1.060 eV (Figure \ref{Fig1}a).

\onecolumngrid

\twocolumngrid

The same energy barrier value is obtained by our ML-NEB method after only 11 function calls (see Fig. \ref{Fig1}b).
In Fig. \ref{Fig1}b we illustrate the evolution of the predicted PES and energy profile along the reaction coordinate (red circles) from the IS to the FS obtained after 1, 3, 10 and 11 iterations of our surrogate machine learning model.
Our algorithm starts by evaluating an image along the initial interpolated path that is located at one third distance from the maximum energy point predicted.
This prevents numerical problems during the optimization of the NEB due to a poor initial representation of the predicted PES when the model is trained with only the two end-points of the transition.
The model is retrained with the energy and forces of the previously evaluated configurations each time a function evaluation is performed.
After training the model, the initial path is optimized on the predicted PES using a velocity-Verlet molecular dynamics algorithm (as implemented in ASE).
Once the elastic band is converged, the energy and uncertainty estimate (blue bars in Fig. \ref{Fig1}b) for each image along the path are stored.
On the basis of these predicted values, an acquisition function suggests the next structure to evaluate (see white circles in Fig. \ref{Fig1}b). 
In this example, the acquisition function targets the image along the predicted path with maximum uncertainty until the uncertainty of all the images ($max|u_{i}|$) is decreased below 0.05~eV.
Once this uncertainty convergence criterion is reached, the acquisition function targets the highest energy image (including the uncertainty estimate), until the maximum force of all the relaxed atoms for the last evaluated image goes below the convergence criteria ($max|$$f_{i}|$$<$0.05~eV/\AA).
This ensures that the saddle-point is obtained with the same accuracy as the classical CI-NEB method.

\begin{figure}[h]
\includegraphics[scale=0.96]{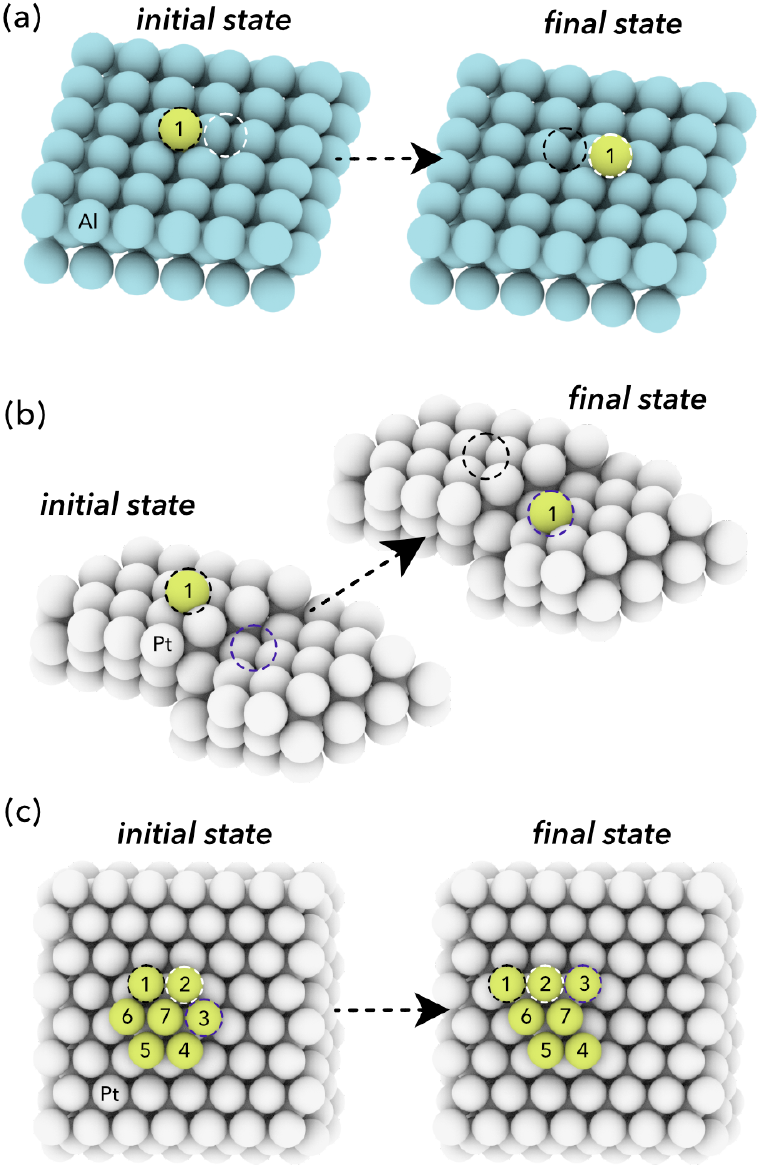}
\caption{Schematic representation of the atomic structures for the initial and final states used for the NEB benchmark calculations: (a) diffusion of a Au atom on an Al(111) surface, (b) diffusion of a Pt adatom on a stepped Pt surface across the two terraces and (c) rearrangement of a Pt heptamer island adsorbed on a Pt(111) surface.}
\label{Fig2}
\end{figure}

We demonstrate the performance of our algorithm on three different atomic systems (see Fig. \ref{Fig2}a-c) using the Effective Medium Theory (EMT) \cite{jacobsen1987interatomic}.
We apply our algorithm using three different acquisition functions:
The first (Acq. 1) alternates between evaluating the image with the maximum uncertainty and the image with the maximum expected energy value for the transition in each iteration of the surrogate model.
This quasi-random sampling mechanism is performed until both convergence criteria are satisfied ($max|$$u_{i}|$$<$0.05~eV and $max|$$f_{i}|$$<$0.05~eV/\AA).
The second acquisition function (Acq. 2) is as described above for the example in  Figure \ref{Fig1}b.
The last (Acq. 3) is made of a combination of the two previous acquisition functions, behaving in the same as Acq. 2 until the uncertainty convergence criteria is satisfied, and then transitions to Acq. 1 until finding a saddle-point.

\begin{figure}[!h]
\includegraphics[scale=0.93]{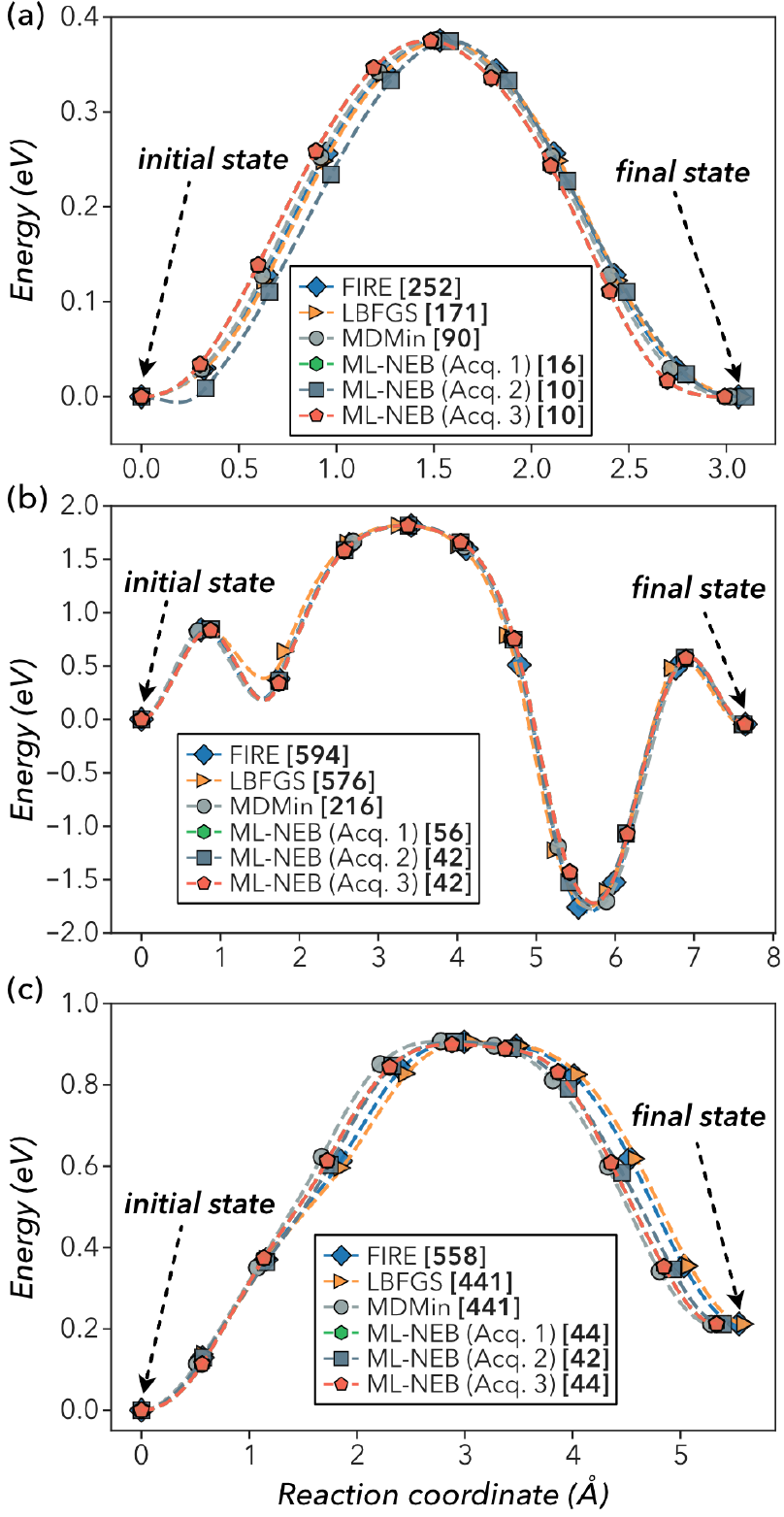}
\caption{Potential energy profiles for the NEBs corresponding to the three transitions schematically shown in Figure \ref{Fig2} (a) to (c), respectively, for the algorithms:  FIRE, LBFGS, MDMin, and ML-NEB (using the three acquisition functions presented in the main text). The number of function calls required for each algorithm to converge are shown in bold between brackets.}
\label{Fig3}
\end{figure}

In Fig. \ref{Fig3} we show the optimized paths for the three transitions illustrated in Fig. \ref{Fig2} using FIRE \cite{bitzek2006structural}, LBFGS \cite{liu1989limited}, and MDMin \cite{larsen2017atomic} as implemented in ASE, along with the ML-NEB implementation using the three acquisition functions described above.
The different algorithms provide virtually identical estimates of the maximum transition state energy.
The same energy barrier values are also obtained when using the classical and ML-NEB algorithms, within numerical precision.
The ML-NEB method performs consistently better in terms of function evaluations than the classical algorithms. 
In particular, when using Acq. 2, the ML-NEB algorithm requires approximately 5-25 times fewer function calls to achieve convergence than the classical algorithms (see values in brackets in Fig. \ref{Fig3}).
The improved performance of the acquisition function which makes the most use of the uncertainty also illustrates the potential for GP to accelerate the NEB over other machine learning algorithms.

\onecolumngrid

\begin{figure}[!h]
\includegraphics[width=0.98\linewidth]{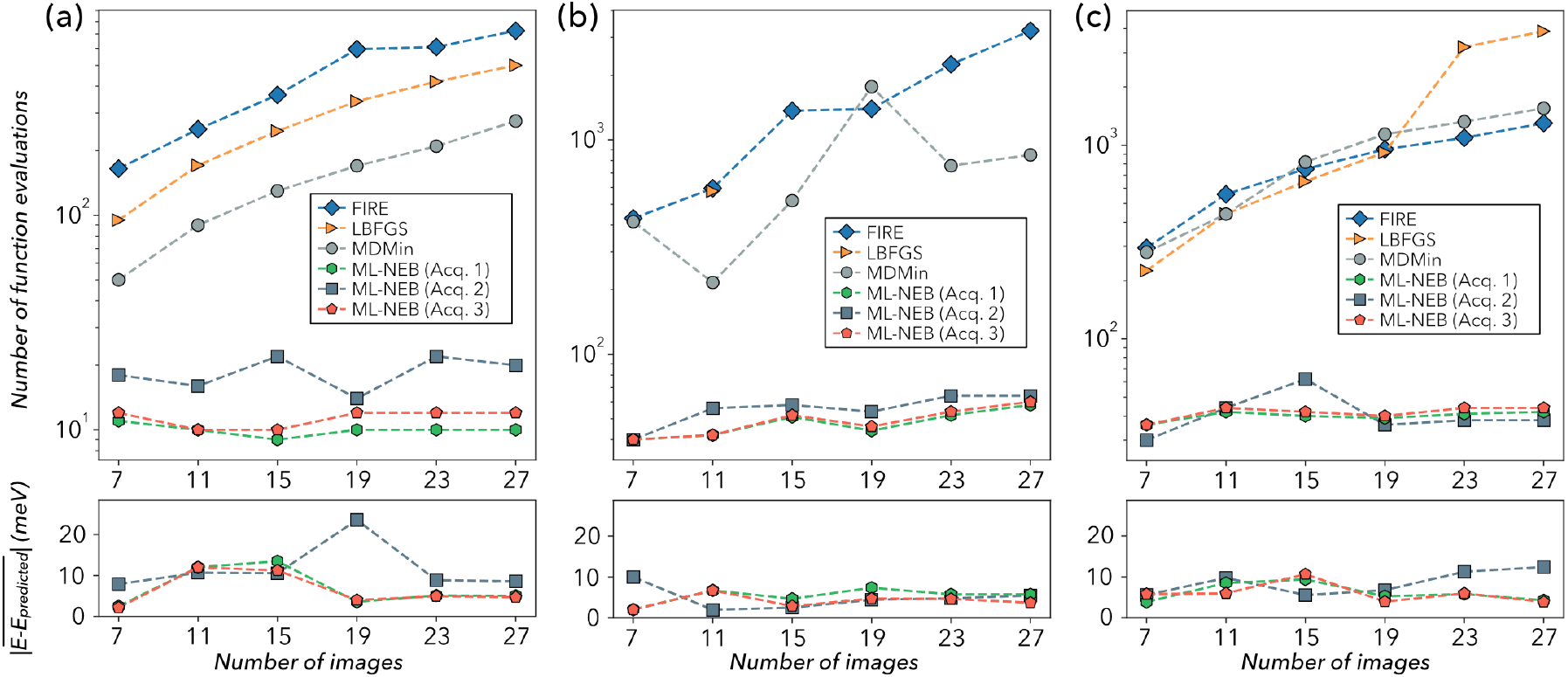}
\caption{Comparison of the number of function evaluations required to achieve convergence with increasing number of images for the different classical and machine learning accelerated methods. The benchmark is performed with the classical method (using the FIRE, LBFGS and MDMin algorithms) and the ML-NEB method (using the three acquisition functions described in the main text).
The lower panels show the average error of the predicted energy along the path obtained by the three acquisition functions with respect to the analytic value of the function at the same geometric positions as the ones predicted by the ML-NEB.}
\label{Fig4}
\end{figure}

\twocolumngrid

The performance of the ML-NEB method is also tested on the previous systems by varying the number of NEB images (see Fig. \ref{Fig4}). 
The number of function calls required to optimize the paths increases exponentially when using the classical implementation of the CI-NEB method.
In contrast, the number of function evaluations required by the ML-NEB algorithm is independent of the number of moving images chosen to optimize the path.
This allows for the number of images to be optimally chosen whilst performing the NEB optimization at no added cost and can be done by applying similar principles to those proposed by Hammer \textit{et al.} \cite{kolsbjerg2016automated} for the classical NEB.\\

In order to quantify the error magnitude of the GPR estimates, we calculated the energy of the predicted images on the analytic potential (EMT) using the same geometries as the images along the optimized path. 
We define the average error of each path as the absolute value of the difference between the energy calculated analytically and GPR predicted energy for the $i$th image along the predicted path.
For the three acquisition functions, the maximum error of the predictions lies below the uncertainty convergence criteria imposed (0.05~eV).
The two acquisition functions that exploit the maximum uncertainty estimate before targeting the saddle-point, Acq. 2 and 3, performed better than Acq. 1 which alternates targets between the maximum energy and the maximum uncertainty estimates in terms of function evaluations and the accuracy of the predicted path.

For stability, the calculations performed using FIRE, MDMin and ML-NEB converged for all three systems.
However, we note that the LBFGS algorithm seems to struggle to find an optimal minimum for the transition represented in Fig. \ref{Fig2}b, except when using 11 images.
We have also encountered convergence issues with MDMin when performing Density Functional Theory (DFT) calculations for validation (see Supplemental Material).
Our algorithm has also been tested on more complex reactions involving bond breaking/forming using DFT \cite{hohenberg1964inhomogeneous, kohn1965self} as implemented in VASP, also included in the Supplemental Material.
Through this variety of examples, our ML-NEB method shows great improvement with respect to the classical optimization in terms of robustness, accuracy, and computational cost. \\
 
A good description of a NEB path ultimately relies on including a sufficient number of images.
Trying to describe the MEP with a small number of images can lead to convergence problems when optimizing the band on complex energy landscapes \cite{sheppard2011paths}.
Here, we have presented a machine learning surrogate model that uses the GPR estimates to obtain a converged NEB path which is independent of the number of moving images composing the path.
This offers a dramatic improvement in terms of the robustness and efficiency with respect to the classical NEB methods.

In this work, we propose three different acquisition functions in an effort to optimize the decision making protocol in order to obtain an accurate predicted path using the smallest possible number of function calls.
We show that the learning rate is driven by the form of the acquisition function and a good selection is dependent on a balance between exploration (reducing the uncertainty of the predicted path) and exploitation (trying to converge the saddle-point).
The result of this work is an algorithm which not only surpasses existing methods in saving function calls, but also improves the robustness in converging an accurate path with respect to the other algorithms, by decoupling the cost in number of function evaluations from the number of moving images on the NEB.

{\em Acknowledgment:} This work was supported by the U.S. Department of Energy, Chemical Sciences, Geosciences, and Biosciences (CSGB) Division of the Office of Basic Energy Sciences, via Grant DE-AC02-76SF00515 to the SUNCAT Center for Interface Science and Catalysis.

\bibliography{literature}

\end{document}


\preprint{APS/123-QED}

\title{Supplemental Material for Low-Scaling Algorithm for Nudged Elastic Band Calculations Using a Surrogate Machine Learning Model}
\author{Jos\'e A. Garrido Torres} %
\author{Paul C. Jennings}%
\author{Martin H. Hansen}%
\author{Jacob R. Boes}%
\affiliation{
Stanford University, Department of Chemical Engineering, Stanford, California 94305, USA
}
\affiliation{
SUNCAT Center for Interface Science and Catalysis, Stanford Linear Accelerator Center, Menlo Park, California 94025, USA.
}
\author{Thomas Bligaard}%
 \email{bligaard@stanford.edu}
\affiliation{
SUNCAT Center for Interface Science and Catalysis, Stanford Linear Accelerator Center, Menlo Park, California 94025, USA.
}

\date{\today}
\maketitle

\section{\label{subsec:theory} Benchmark using Density Functional Theory}
\begin{figure}[h!]
\includegraphics[scale=0.77]{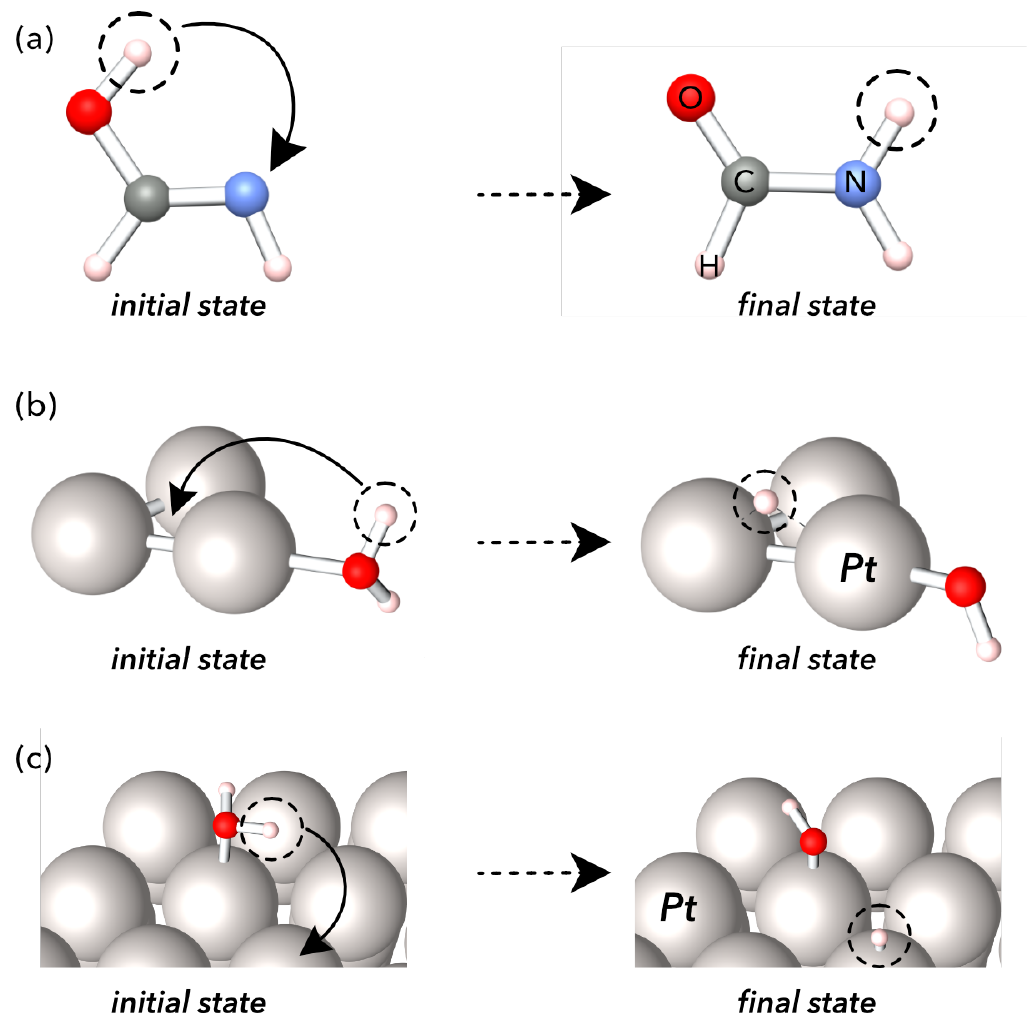}
\caption{Schematic representation of the atomic structures for the initial and final states used for the NEB benchmark calculations using DFT: (a) keto-enol tautomerization of formamide (HCONH$_2$), (b) water dissociation on a Pt$_3$ cluster and (c) water dissociation on a Pt(111) surface.}
\label{FigS1}
\end{figure}

In this benchmark we tested the performance of the different algorithms (number of function evaluations required to converge the NEB path) using Denisty Functional Theory (DFT) for three different systems:
(a) keto-enol tautomerization of formamide, (b) dissociation of H$_2$O on a Pt$_3$ cluster and (c) dissociation of H$_2$O on a Pt(111) surface. The images for the initial and final states for each transition are represented in Fig. \ref{FigS1}.
We used 11 images to describe the reaction path between the initial and final states.
For ML-NEB, we use the same Gaussian Process Regressor (GPR) parameters as the EMT calculations shown in the main text.

\begin{table}[h!]
\centering
\caption{DFT Benchmark. Number of function evaluations required by the FIRE, BFGS and ML-NEB (including the three acquisition functions explained in the main text) to converge the NEB paths describing the following reactions: keto-enol tautomerization of formamide (HCONH$_2$), water dissociation on a Pt$_3$ cluster (H$_2$O/Pt$_3$) and water dissociation on a Pt(111) surface (H$_2$O/Pt(111)).}
\begin{ruledtabular}
\begin{tabular}{lccc}
& HCONH$_2$ & H$_2$O/Pt$_3$ & H$_2$O/Pt(111) \\
\hline     
FIRE & 603 & 1116 & 1134 \\
BFGS &  306 & 855  & 738 \\            
ML-NEB (Acq.1) & 26 & 84 & 106 \\
ML-NEB (Acq.2) & 24 & 61 & 48 \\
ML-NEB (Acq.3) & 28 & 66 & 64 \\
\end{tabular}
\end{ruledtabular}
\label{TableS1}
\end{table}

For these examples, the FIRE performs the worst, followed by the BFGS algorithm. 
This is the same trend observed for the examples in the main text.
Each requires between 10-46 times more function evaluations than the ML-NEB algorithm.
The MDMin algorithm is not included in this benchmark due to convergence problems.
Again, the ML-NEB (Acq. 2) from the main text requires the fewest function evaluations.
These extended examples demonstrate that our method is robust and can be extended to a large number of systems for accelerating transition state search calculations using DFT.

\section{\label{subsec:theory} Computational details}
The periodic DFT calculations were performed using PAW (Projector Augmented Wave \cite{blochl1994projector}) as implemented in VASP \cite{kresse1996efficient, kresse1999ultrasoft}. Valence electrons were described using plane-waves considering an expansion on the kinetic energy up to an energy cut-off of 400~eV. The electron population distribution is integrated using a Gaussian smearing with width of 0.1 eV.   
The calculations performed using the PBE \cite{perdew1996generalized} functional including Grimme's D3(BJ) \cite{grimme2010consistent, grimme2011effect} dispersion correction scheme.
The convergence criterion for the electronic self-consistent cycle was set to 10$^{-5}$ eV.
Convergence on the forces on all ions were required to be smaller than 0.02~eV/\AA~ and 0.05 ~eV/\AA~ for the structure relaxations and NEB calculations, respectively.  

The metal$-$metal distance of the Pt$_3$ cluster was optimized during the geometry relaxation of the initial and final configurations for the transition, and then the Pt atoms were kept fixed during the NEB calculations.
The Pt(111) surface was modeled using a (3$\times$3) slab composed of three layers with the top layer relaxed during optimization. A vacuum of $\sim$10~\AA\ was introduced in the direction orthogonal to the surface. 
The integration of the Brillouin zone for the calculations involving this slab was performed using 2$\times$2$\times$1 \textit{k}-points, whilst the $\Gamma$-point was used for the gas phase calculations.
We use a 10$\times$10$\times$10 \AA$^3$ unit cell for describing the keto-enol tautomerization of formamide.
Water dissociation on the Pt$_3$ cluster was modelled on a 15$\times$15$\times$15 \AA$^3$ simulation cell.
In both cases the vacuum selected was enough to avoid spurious interactions between periodic images.

\bibliography{literature}